# IMPi: An interactive homology modeling pipeline

Rowan Hatherley, David K. Brown and Özlem Tastan Bishop[*]

Research Unit in Bioinformatics (RUBi), Department of Biochemistry and Microbiology, Rhodes University, Grahamstown, 6140, South Africa

*To whom correspondence should be addressed.

## Abstract

**Summary:** The development of automated servers to predict the three-dimensional structure of proteins has seen much progress over the years. These servers make modeling simpler, but largely exclude users from the process. We present an Interactive Modeling Pipeline (IMPi) for homology modeling. The pipeline simplifies the modeling process and reduces the workload required by the user, while still allowing engagement from the user during every step. Default parameters are given for each step, which can either be modified or supplemented with additional external input. As such, it has been designed for users of varying levels of experience with homology modeling.

**Availability and implementation:** The IMPi site is free for non-commercial use and can be accessed at https://impi.rubi.ru.ac.za.

**Supplementary information:** Documentation available at https://impi.rubi.ru.ac.za/#documentation.

**Contact:** O.TastanBishop@ru.ac.za

## 1 Introduction

Computational methods to predict the three-dimensional structures of proteins have become increasingly important in bridging the ever-expanding gap between protein sequence and structural databases (Haas *et al.*, 2013). Homology modeling is a widely-used approach that models proteins based on homologs with known structures (templates). The technique works as protein structures are far more conserved than their sequences, and has been shown to produce accurate models if a suitable template structure is available (Zhang, 2009).

Homology modeling is a multi-step process which, for non-expert users, can be quite daunting; thus many turn to automated online modeling servers. While many fully-automated servers have achieved promising accuracy (Huang *et al.*, 2014), they are in effect a black-box since users are almost entirely excluded from the modeling process. This prevents novice users from learning how to model proteins, and limits the control that the more advanced users may have over their modeling jobs.

To address this problem, we have developed the Interactive Modeling Pipeline (IMPi). It has been designed to guide users through each stage in the homology modeling process. Keeping novice users in mind, the interface is simple and easy to learn, while allowing more experienced users to alter parameters and have control over their modeling jobs.

## 2 Website overview

*Input page:* This provides an overview of the modeling job. Users can simply enter a target sequence, and begin the modeling process. IMPi utilizes MODELLER (Sali and Blundell, 1993), so users must also supply a MODELLER key. After entering a target sequence, users can choose whether to search for templates using HHPred (Söding *et al.*, 2005), or specify templates themselves. They may also select one of five sequence alignment options available (described in the Supplementary Material) and adjust modeling parameters before the modeling job is started. Thereafter, the IMPi interface guides users through each step in the homology modeling process, and gives suggestions to help novice users at each stage. Input for each stage is processed and submitted to our local cluster, utilizing the Job Management System (JMS) (Brown *et al.*, 2015), as explained in the Supplementary Material.

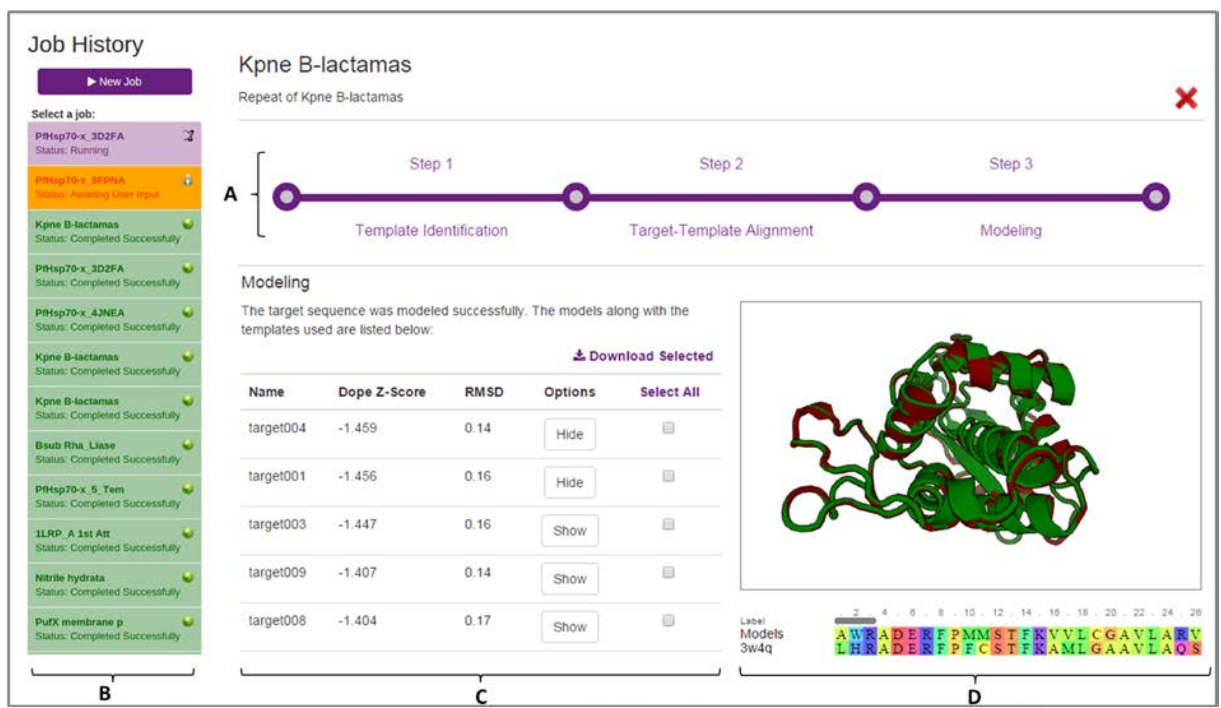

**Fig. 1: Example of the modeling results page.** Results are shown for modeling β-lactamase from *Klebsiella pneumoniae*. The progress bar (A) and job history set (B) allow navigation within current jobs and between different modeling jobs. Jobs are shown as completed (green), awaiting user input (yellow) and running (purple). Models are tabulated (C), ranked by their DOPE Z-scores. This table can be used to select and download models produced, as well as show them in the interactive protein viewer (D). The top model (green) is shown superimposed with the template used to model it (red). Additional details about the protein viewer are provided in the Supplementary Material.

*Template identification:* If HHPred is run, the templates identified are displayed, including information about sequence identity and query coverage. Templates can be selected through simple check boxes to be included in the target-template alignment stage. Users can also click on the ID of any template, which links directly to its entry in the PDB, allowing users to further assess the quality of each template.

*Target-template alignment:* Sequences are extracted from the templates and aligned to the target sequence, using the alignment option selected. The alignment is displayed in an integrated alignment viewer and can be inspected and edited manually by the user before moving on to the modeling stage.

*Model building and evaluation:* The sequence alignment is used to prepare a PIR file as explained in the Supplementary Material. Modeling is performed using the parameters specified in the input page and the models are assessed and ranked by DOPE Z-score calculations. All models can be viewed using the integrated PDB viewer provided (Figure 1).

*Job history:* Jobs are linked to the user account, which provides an instant sign-in. Users can navigate to previous jobs run, as well as to different stages in their current jobs, alter parameters and rerun jobs.

## 3 Conclusion and outlook

As a modeling tool, IMPi aims to provide a middle ground between the lack of control caused by full automation and the difficulty of writing scripts. As a web interface, IMPi is platform independent, and requires no local computing power. The site currently provides a means for modeling protein monomers using one or more templates. Future development will focus on providing more features, such as protein-protein and protein-ligand complex modeling. IMPi encourages user involvement in the homology modeling process and consequently we also aim to provide additional options for each of the stages.

## Acknowledgements

The authors wish to acknowledge members of the Research Unit in Bioinformatics (RUBi) for testing the site and providing valuable input to improve IMPi.

## Funding

This work was supported by the National Institutes of Health Common Fund [grant number U41HG006941] to H3ABioNet; the National Research Foundation (NRF), South Africa [grant numbers 79765, 93690]; and Rhodes University, Postdoctoral Fellowship. The content of this publication is solely the responsibility of the authors and does not necessarily represent the official views of the funders.

*Conflict of Interest:* none declared.

# Supplementary information for "IMPi: An interactive homology modeling pipeline"

Rowan Hatherley, David K. Brown and Özlem Tastan Bishop[*]

Research Unit in Bioinformatics (RUBi), Department of Biochemistry and Microbiology, Rhodes University, Grahamstown, 6140, South Africa

*To whom correspondence should be addressed.

## 1   Submitting jobs to the cluster via JMS

IMPi makes use of a unique system to submit jobs to the underlying cluster. The Job Management System (Brown *et al.*, 2015) (JMS) has been developed as a web-based workflow management system and cluster front-end for high performance computing (HPC). It is able to store custom tools and scripts and manage their execution on an HPC cluster. The reason that JMS was used for submitting jobs was that it abstracted away the complexity of managing the job on the cluster and drastically reduced the time taken to develop the IMPi web server. IMPi was originally developed as a series of command-line scripts. We were able to upload these scripts to JMS directly via the JMS web interface. After that, building the IMPi website simply involved building a custom interface that interacted with the JMS web API. Submitting and managing the job on the cluster was handled entirely by JMS while the IMPi web server merely had to wait for a notification from JMS that the job had completed.

The diagram presented in Fig. S1 illustrates the process by which jobs are submitted from IMPi to the cluster via JMS. When a user submits a modeling job from the IMPi interface, their input parameters are sent to the IMPi server. IMPi then compiles these parameters into a request to be sent to JMS. Authentication details for JMS are also added to the request at this point. Once the request has been compiled, it is sent to JMS, which submits the job to the cluster and returns the job ID to the IMPi web server. IMPi then simply returns a message to the interface that the job was submitted successfully, while JMS monitors the job on the cluster. When the job finishes running on the cluster, JMS notifies IMPi that the results are available. IMPi then collects the results and returns them to the interface, where the user can interact with them.

## 2   Sequence alignment options

There are five alignment options made available when using IMPi. The first four include MAFFT (Katoh et al., 2002), MUSCLE (Edgar, 2004), T-COFFEE (Notredame et al., 2000) and Clustal-OMEGA (Sievers et al., 2011). An additional option involves using the alignment produced when searching for templates with HHPred (Söding et al., 2005). This option may only be selected if HHPred is used in the template identification step.

## 3   PIR file preparation

The software for IMPi consists of a series of Python classes, with all integrated into a single overall application, designed to handle the stages presented on the interface. Two of these classes in particular add to the ease with which IMPi can be used: these are the PIR module and the PDB parser class. Together these classes perform the PIR file preparation, which takes place before modeling. The PDB parser reads PDB files and extracts useful information in a format that can be used throughout the IMPi application and is essential to the PIR module.

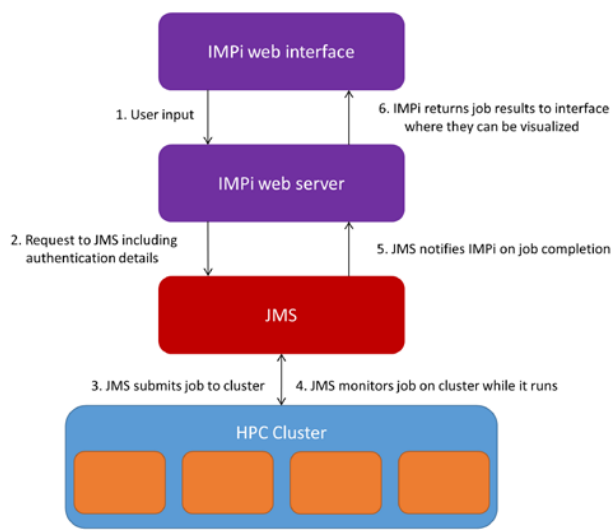

**Fig. S1: Submitting jobs via JMS.** The above figure illustrates the process of submitting a job to the cluster via JMS

*R.Hatherley et al.*

A PIR file maps each template sequence to their location with their PDB files, from starting to ending residues and on which chain, while maintaining their alignment to the target sequence. The PIR file also links the modelling Python script to the template PDB files. The names of each template PDB file are provided in the PIR file such that MODELLER can identify and use them. This naming, as well as the residue numbering needs to be done correctly in order for MODELLER to run without crashing.

During PIR file preparation, sequences are extracted from the alignment step and converted to PIR file format. Each template sequence is then inspected, alongside information from the PDB parser. Firstly, If any missing residues are present in the sequence, they are removed and modified residues are converted to a '.' character, so as to be read by MODELLER as such. Both the start and ending residues of each sequence are determined and included in the PIR header, along with the chain information already specified.

A trimming phase is also included, which removes overhanging sequence segments at both the N- and C-termini of the target sequence. If these are left in, MODELLER will attempt to model them, but with poor results in the absence of any template information.

In our experience, PIR file preparation is the most challenging aspect when learning to use MODELLER, as it is easy for a user to make a number of different mistakes in this step, but difficult to discern exactly what they did wrong. The PIR file preparation step included in IMPi minimizes unnecessary errors, allowing users to focus on adjusting parameters of their modeling jobs, rather than wasting time on initial set up.

## 4 PV-MSA: a JavaScript wrapper combining the functionality of PV and MSA

PV (Biasini, 2015) is a widely used JavaScript plugin for three dimensional protein visualization. Similarly, BioJS MSA is a JavaScript component used to visualize multiple sequence alignments. Although useful in their own right, the need to view a structure in conjunction with its sequence often arises in bioinformatics. In addition, these tools can be difficult to use as their application programming interfaces (API) are fairly unintuitive. To cater for this, we have developed PV-MSA, a wrapper that combines the functionality of PV and MSA in a single JavaScript plugin. PV-MSA also provides a simplified API that makes a fair amount of the functionality of both PV and MSA available. For functionality that has not been included yet, PV-MSA provides direct access to the underlying PV and MSA objects.

Over and above simply wrapping the two plugins, PV-MSA links the selection functionality. For example, a user can select a residue in the protein structure and it will automatically be highlighted in the alignment. Similarly, if a residue is selected in the alignment, its location is highlighted on the corresponding structure. PV-MSA, of course, also allows structures to be superposed. In such cases, selecting a residue in one structure will also highlight the aligned residue in the superposed structure. This selection is based off of the alignment, and as such, gaps and missing residues are taken into account.

Multiple structures and their sequences can be loaded into the plugin at once and structures and sequences can be hidden and shown independently. Functionality has also been included to resize both the PV and MSA plugins together and independently as the user needs.

The PV-MSA interface is illustrated in Fig. S2. The PV-MSA plugin can be downloaded from https://github.com/davidbrownza/PV-MSA.

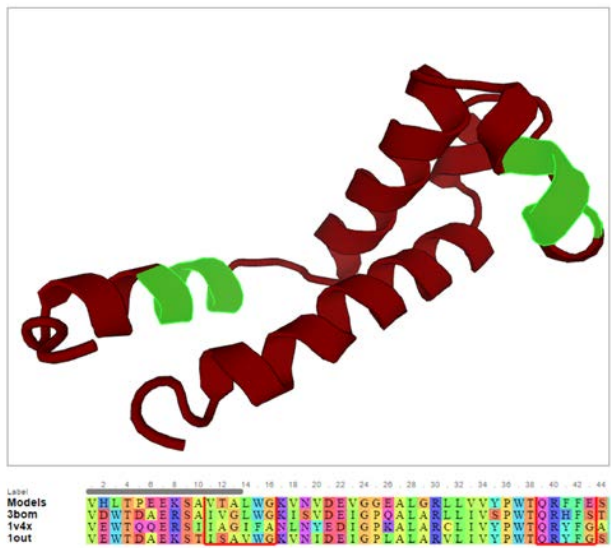

**Fig. S2: The PV-MSA plugin interface.** By default, the alignment is displayed underneath the structure. Selecting residues in the structure highlights them in bright green and puts a red box around the corresponding columns in the alignment and vice versa.